\documentclass[apjl, iop]{emulateapj}
\shorttitle{An extremely luminous radio-loud quasar at $z=5.18$}
\shortauthors{Yi et al.}

\begin{document}


\title{SDSS J013127.34$-$032100.1: A newly discovered radio-loud quasar at $z=5.18$ with extremely high luminosity }


\author{
  Wei-Min Yi$^{1,2,3}$,
  Feige Wang$^{4}$,
  Xue-Bing Wu$^{4,5}$,
  Jinyi Yang$^{4}$,  
  Jin-Ming Bai$^{1,3}$,
  Xiaohui Fan$^{6}$, 
  William N. Brandt$^{7}$, 
  Luis C. Ho$^{5,4}$,
  Wenwen Zuo$^{4,8}$, 
  Minjin Kim$^{9}$,
  Ran Wang$^{5}$,
  Qian Yang$^{4}$,  
  Ju-jia Zhang$^{1,2,3}$,
  Fang Wang$^{1,2,3}$,
  Jian-Guo Wang$^{1,3}$,
  Yanli Ai$^{4}$,
  Yu-Feng Fan$^{1,3}$,
  Liang Chang$^{1,3}$,
  Chuan-Jun Wang$^{1,2,3}$,
  Bao-Li Lun$^{1,3}$,
  Yu-Xin Xin$^{1,3}$
}


\altaffiltext{1}{Yunnan Observatories, Chinese Academy of Sciences, Kunming 650011, China.}
\altaffiltext{2}{University of the Chinese Academy of Sciences, Beijing 100049, China}
\altaffiltext{3}{Key Laboratory for the Structure and Evolution of Celestial Objects, Chinese Academy of Sciences, Kunming 650011, China.}
\altaffiltext{4}{Department of Astronomy, School of Physics, Peking University, Beijing 100871, China}
\altaffiltext{5}{Kavli Institute for Astronomy and Astrophysics, Peking University, Beijing 100871, China}
\altaffiltext{6}{Steward Observatory, University of Arizona, Tucson, AZ 85721-0065, USA}
\altaffiltext{7}{Department of Astronomy \& Astrophysics, The Pennsylvania State University, 525 Davey Lab, University Park, PA 16802, USA}
\altaffiltext{8}{Shanghai Astronomical Observatory, Chinese Academy of Sciences, Shanghai 200030, China}
\altaffiltext{9}{Korea Astronomy and Space Science Institute, Daejeon 305-348, Korea}


\begin{abstract}
Only very few $z>$5 quasars discovered to date are radio-loud, with a radio-to-optical flux ratio (radio-loudness parameter) higher than 10. Here we report the discovery of an optically luminous radio-loud quasar, SDSS J013127.34$-$032100.1 (J0131$-$0321 in short), at $z=5.18\pm$0.01 using the Lijiang 2.4m and Magellan telescopes. J0131$-$0321 has a spectral energy distribution consistent with that of radio-loud quasars. With an $i$-band magnitude of 18.47 and radio flux density of 33 mJy, its radio-loudness parameter is $\sim$ 100. The optical and near-infrared spectra taken by Magellan enable us to estimate its bolometric luminosity to be $ L_{\rm bol} \sim1.1 \times10^{48}$ erg s$^{-1}$, approximately 4.5 times greater than that of the most distant quasar known to date. The black hole mass of J0131$-$0321 is estimated to be $2.7\times10^9 M_\odot$, with an uncertainty up to 0.4 dex. Detailed physical properties of this high-redshift, radio-loud, potentially super-Eddington quasar can be probed in the future with more dedicated and intensive follow-up observations using multi-wavelength facilities.
\end{abstract}





\section{Introduction}

Identifying a sample of high-redshift, radio-loud quasars provides important clues about the early Universe and potentially can probe formation of massive galaxies. \citet{Jiang2007a} found that the radio-loud fraction of quasars is a strong function of both redshift and optical luminosity, such that  the radio-loud fraction decreases rapidly with increasing redshift and decreasing luminosity. Although the expected number of radio-loud sources with redshift smaller than 3 is fairly consistent with the optically selected quasars from the Sloan Digital Sky Survey (SDSS), when it comes to high-redshift cases, there is an apparent deficit of SDSS radio-loud quasars \citep{Volonteri}. It still remains to be seen whether the small number of high-redshift radio-loud quasars discovered to date is due to selection effects owing to current limitations of the high redshift identification techniques, or whether these quasars are intrinsically rare. Radio-emitting, high-redshift objects may be particularly valuable in addressing questions with respect to early cosmological evolution. These rare, distant, and powerful objects provide important insights into the activity of supermassive black holes (BHs) at early cosmological epochs and into the physical properties of their environments. Rapid galaxy formation and starburst activity may go hand-in-hand with the radio-active phase of accreting supermassive BHs \citep{Frey2010}.

Recent studies (e.g., \citealp{Barth,Jiang2007b,Kurk,Willott}) show that high-redshift quasars with $M_{\rm BH}$ $>$ 10$^9 M_\odot$  are accreting close to the Eddington limit ($L_{\rm bol}$ /$L_{\rm Edd}$ $\approx$ 1), while at lower redshifts ($2 < z < 3$) the most luminous quasars ($L_{\rm bol}$ $>$ 10$^{47}$ erg s$^{-1}$) are characterized by $L_{\rm bol}$/$L_{\rm Edd}$ $\approx$ 0.25 with a dispersion of 0.23 dex (e.g., \citealp{Shen2008}). Furthermore, \citet{De Rosa} found that the distribution of observed Eddington ratios for high-redshift quasars is significantly different than that of a luminosity-matched comparison sample of SDSS quasars at lower redshift ($0.35 < z < 2.25$): the $z>4$ sources are accreting significantly faster than the lower redshift ones, which implies that the masses of these BHs at high redshift are typically lower than those of the lower redshift quasars. High-redshift quasars, especially highly radio-loud ones, may have intrinsically different properties compared to lower redshift ones.

We report the discovery and first observational results of an extremely luminous and radio-loud quasar, SDSS J0131$-$0321, at $z=5.18$. It appears similar to another radio-loud source, SDSS J0741+2520 \citep{McGreer}, at nearly the same redshift, but J0131$-$0321 is more luminous at optical and radio wavelengths. The high luminosity of J0131$-$0321 may be due to super-Eddington accretion. Throughout this Letter, luminosity distances are calculated using a $\Lambda$-dominated flat cosmology with $H_0$ = 70 km s$^{-1}$Mpc$^{-1}$,  $\Omega_m$ = 0.3 and  $\Omega_{\Lambda}$ = 0.7.

\section{Observations}

%% In a manner similar to \objectname authors can provide links to dataset
%% hosted at participating data centers via the \dataset{} command.  The
%% second curly bracket argument is printed in the text while the first
%% parentheses argument serves as the valid data set identifier.  Large
%% lists of data set are best provided in a table (see Table 3 for an example).
%% Valid data set identifiers should be obtained from the data center that
%% is currently hosting the data.
%%
%% Note that AASTeX interprets everything between the curly braces in the 
%% macro as regular text, so any special characters, e.g. "#" or "_," must be 
%% preceded by a backslash. Otherwise, you will get a LaTeX error when you 
%% compile your manuscript.  Special characters do not 
%% need to be escaped in the optional, square-bracket argument.

J0131$-$0321 was first selected as a candidate high-redshift quasar according to optical-IR selection criteria based on SDSS and {\it WISE}\ photometric data \citep{wxb}.  The first low-resolution optical spectrum of this source was obtained with the Lijiang 2.4m telescope (LJT, see \citealt{Fan}) using the Yunnan Fainter Object Spectrograph and Camera (YFOSC, \citealt{Zhang}) on November 25, 2013 (UT).  With a 25 min exposure, this spectrum taken with YFOSC\underline~G3 grism ($\lambda / \Delta\lambda \sim$ 680 at 7000\AA~ with a 1.\arcsec 0 slit) in a long-slit mode yields, through the identification of Ly$\alpha$ emission, a redshift close to 5.2. It has a signal-to-noise ratio (S/N) of 8.3 per pixel in the continuum. Due to the relatively low S/N and a cut-off at 9000\AA~ in the first spectrum, another spectrum was obtained with a more red-sensitive grism (YFOSC\underline~G5, $\lambda / \Delta\lambda \approx$ 550 at 7000\AA~ with a 1.\arcsec 0 slit) by the LJT during a photometric night on February 20, 2014 (UT). The average seeing (1\arcsec.2) during the second observation was better than the first.

In order to obtain more reliable estimates of the continuum luminosity and SMBH mass of J0131$-$0321, we obtained subsequent observations in the optical and near-IR bands with higher resolution spectroscopy using the Magellan Echelette (MagE) and FIRE spectrographs on January 3, 16 and 18, 2014 (UT), with 90 and 50 min of integration time, respectively. 
\begin{figure}[h]
\epsscale{1.15}
\plotone{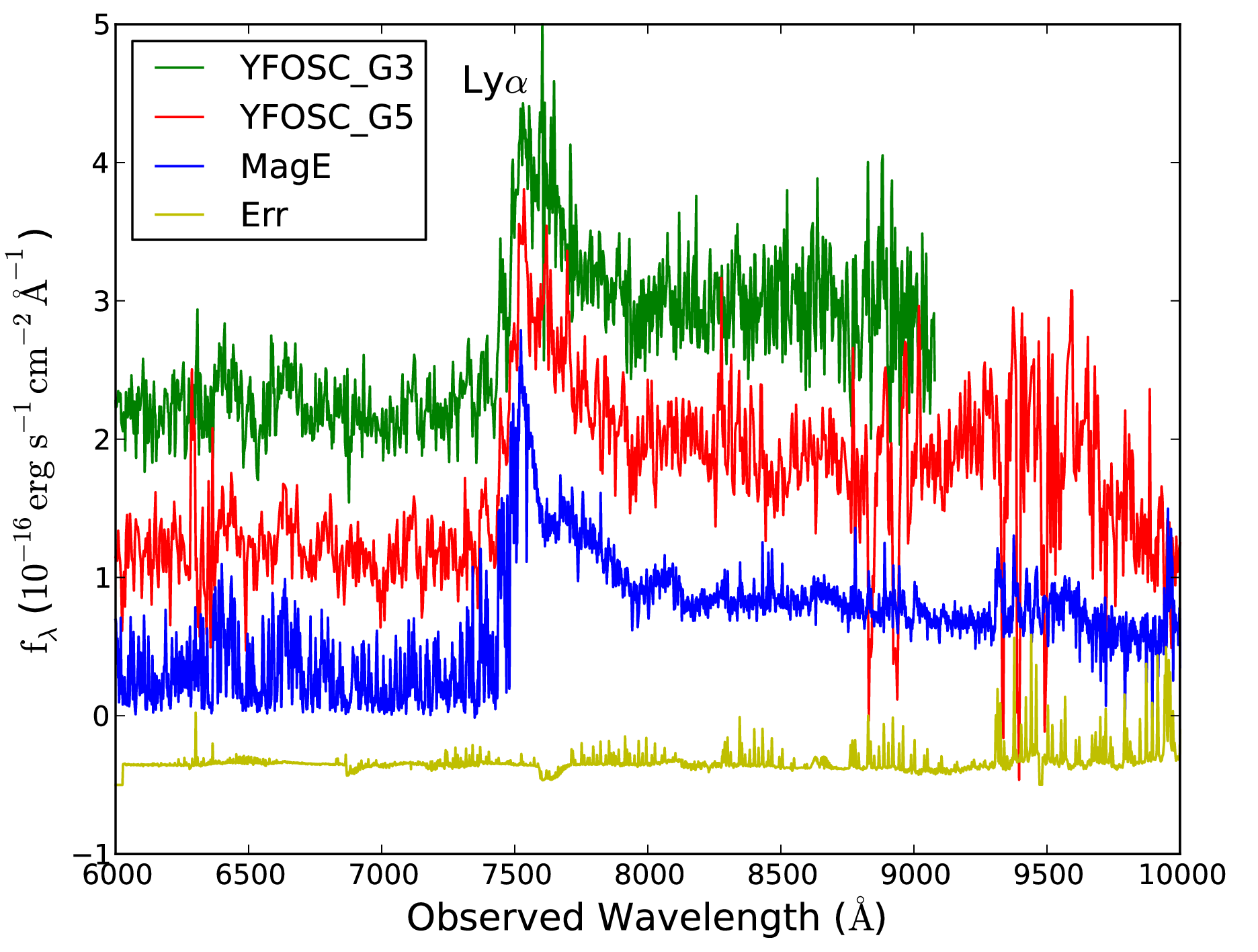}
 \caption{The optical spectra of J0131$-$0321 (telluric corrected and flux calibrated) taken with the YFOSC and Magellan Echelette spectrographs. The yellow curve (bottom one) is the error distribution when extracting the spectrum from MagE raw data, which is shifted downward by a half unit on the y-axis. Two YFOSC spectra taken with Grism 5 and 3 are shifted upward one and two units on the y-axis, respectively.}
  \label{fig1}
\end{figure}
The MagE optical spectrum provides a resolution of $\lambda /\Delta\lambda =4100$ with a $1\arcsec$ slit, and the FIRE spectrum covers 8200 \AA~  to 25000 \AA~  at a spectral resolution of $\lambda /\Delta\lambda = 6000$. These higher resolution spectra lead to a more accurate redshift of z=5.18$\pm$0.01 based on the Ly$\alpha$ and Mg II lines. The raw data are processed in a standard way, including bias, flat correction, sky subtraction, flux calibration and telluric correction. The region of overlap with the optical spectrum is used to calibrate and normalize the near-IR flux more accurately. All the optical spectra plotted in Fig. 1 are scaled to the SDSS photometric magnitude, and the spectrum taken with YFOSC\underline~G5 on a photometric night is considered as a standard to recalibrate the other two spectra.  For the purposes of display and comparison with the YFOSC spectra, the MagE spectrum has been smoothed with a 5-pixel boxcar filter.

Photometric points from radio to UV/optical bands were collected from multiple surveys, including FIRST, 2MASS, {\it WISE}\ and SDSS \citep{Becker,Skrutskie,Wright,York}. The radio flux density of J0131$-$0321 is 33 mJy at 1.4 GHz (observed-frame).  The source appears core-dominated, as no extended structure is visible on the 5$\arcsec$ resolution scale of FIRST.

%% In this section, we use  the \subsection command to set off
%% a subsection.  \footnote is used to insert a footnote to the text.


%% This section also includes several of the displayed math environments
%% mentioned in the Author Guide.

\section{Luminosity Analysis: From UV/Optical to Radio Wavelenths}

\begin{figure}[h]
\epsscale{1.15}
\plotone{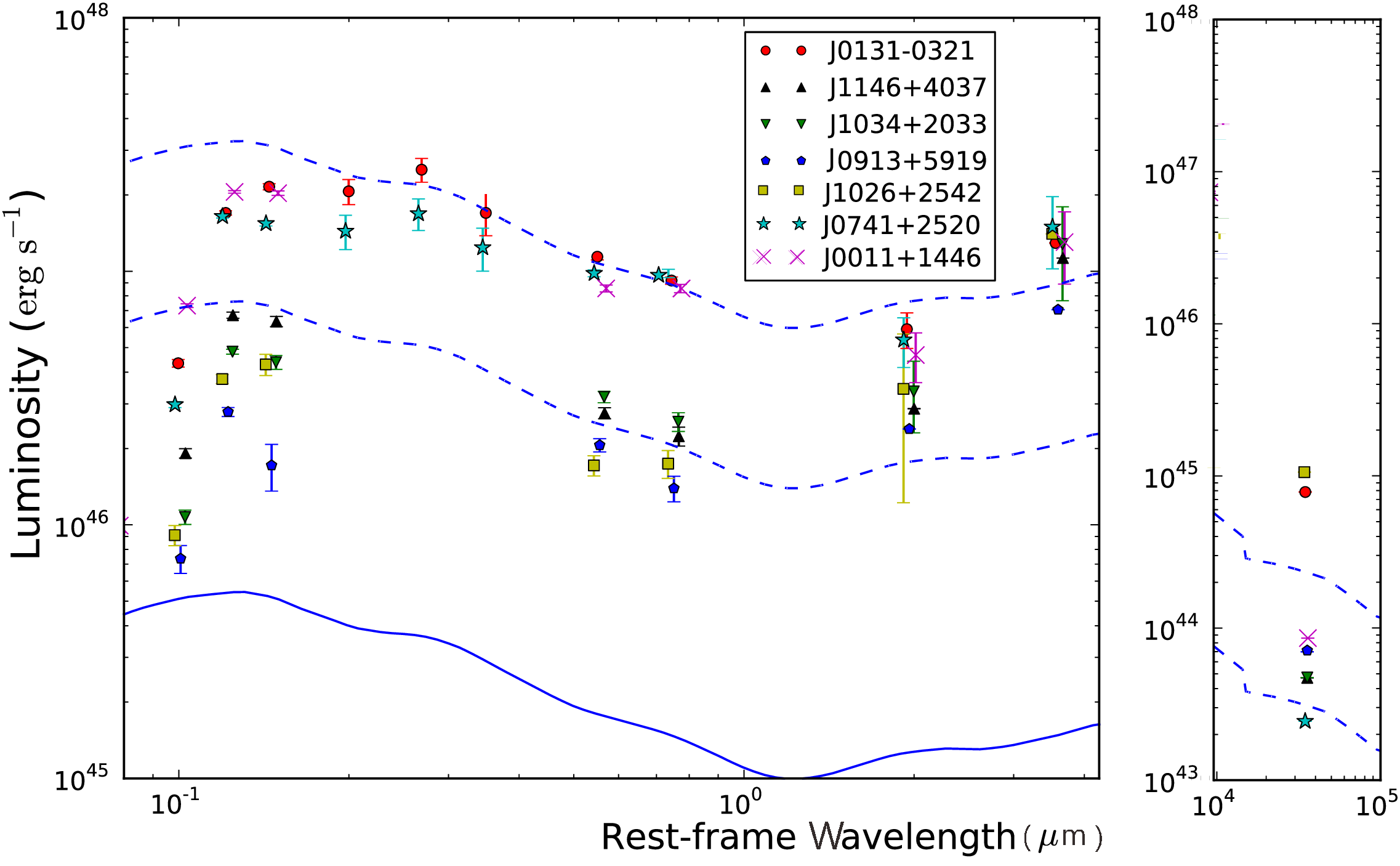}
  \caption{Rest-frame SEDs of seven $z\approx 5$ radio-loud quasars compiled from multiple photometric surveys (FIRST, 2MASS, {\it WISE}\ and SDSS). The data points plotted here are all calibrated in the rest-frame after K-correction and Galactic extinction correction. The approximate rest-frame effective wavelengths of the left and right panels are located between 0.1 and 4 $\mu$m, 10$^4$ and 10$^5 \mu$m, respectively. The blue solid line at the bottom is the mean SED template of radio-loud quasars in the local Universe \citep{Elvis}, which is significantly less luminous than the seven high-redshift radio-loud quasars. Two dashed lines, shifted upward from the blue solid line, are shown to roughly bracket the range of optical luminosities of the high-redshift sample.  Apart from the stronger Ly$\alpha$ absorption and radio emission, the UV/optical spectral indices of the seven high-redshift quasars are similar to that of the low-redshift counterparts.}
\label{fig2}
\end{figure}

\begin{table*}
 \centering
 \begin{minipage}{150mm}
  \caption{7 radio-loud quasars at  $z \approx$ 5}
  \begin{tabular}{@{}lccccrcrc@{}}
  \hline
   Object Name & R.A. (J2000) & Decl. (J2000) &  $z\tablenotemark{a}$ & $m_i\tablenotemark{b}$ & $M_i\tablenotemark{c}$
     & $f_{\rm 1.4GHz}\tablenotemark{d}$ & $logR\tablenotemark{e}$  & Reference$\tablenotemark{f}$  \\
        & (deg) & (deg) &  & (AB mag) &  & (mJy) & & \\
        \\
 \hline
 \\
 J0131$-$0321  & 22.86392 & $-$3.35005 & 5.18 & 18.46 & $-$29.11 & 32.9 & 2.02 &  1,8,9 \\
 J0011+1446 & 2.81349 & 14.76717 & 4.96 & 18.29 & $-$29.01 & 24 & 2.08 & 2,6,7,9,10 \\
 J1146+4037 & 176.74080 & 40.61907 & 5.01 & 19.41 & $-$28.08 & 12.4 & 2.15 & 2,5,9 \\
 J1034+2033 & 158.57771 & 20.55006 & 5.01 & 19.79 & $-$27.76 & 3.96 & 1.66  & 2,4,9 \\
 J0913+5919 & 138.31896 & 59.32269 & 5.12 & 20.46 & $-$27.14 & 17.5 & 2.61  & 2,3,9 \\
 J0741+2520 & 115.47799 & 25.34157 & 5.19 & 18.54 & $-$29.04 & 2.97 & 1.11 & 3,4,9 \\
 J1026+2542 & 156.59845 & 25.71651 & 5.26 & 20.17 & $-$27.64 & 239 & 3.69 & 2,3,4,9 \\
 \\
\hline
\end{tabular}\\
\tablenotemark{a}{Redshift.}  
\tablenotemark{b}{Apparent $i$-band magnitude.}
\tablenotemark{c}{Absolute $i$-band magnitude, corrected for Galactic extinction. }
\tablenotemark{d}{Peak flux density at 1.4 GHz from the FIRST survey \citep{Becker}. }
\tablenotemark{e}{Logarithmic value of the radio-loudness parameter defined by $R=f_{\rm 5GHz}$/$f_{\rm 4400~\AA}$\citep{Kellermann}.}
\tablenotemark{f}{References: (1) this work; (2) \citet{Shen2011}; (3) \citet{Wu};  (4) \citet{McGreer};  (5) \citet{Ghisellini};  (6) \citet{Shemmer};(7) \citet{Shen2012};(8) \citet{Skrutskie}; (9) \citet{Becker};  (10) \citet{Schneider}  }
\end{minipage}
\end{table*}

For simplicity, we use the UV/optical continuum luminosity as a measure of the energy output of the quasar. We corrected the photometric magnitude for Galactic reddening using the $E(B-V)$ values from \citet{Schlafly}. In Fig. 2, the blue solid line is the mean spectral energy distribution (SED) for local radio-loud quasars \citep{Elvis}. The photometric data points of J0131$-$0321 are roughly consistent with the upward-scaled template for local radio-loud quasars  (the top dashed line), except for the stronger Ly$\alpha$ absorption seen in the bluest bands of J0131$-$0321. Six additional high-redshift ($z\approx 5$; see Table 1) radio-loud quasars are shown for comparison.  It is apparent that J0131$-$0321 is the most luminous among them. Compared with other less luminous quasars at $z\approx 5$ tabulated in Table 1, one may easily postulate that these highly luminous quasars are little affected by the UV/optical obscuration along the line of sight or there may exist significant jet contributions or lensing effects. In addition, it is noteworthy that J0011+1446 \citep{Schneider} and J0741+2520 \citep{McGreer} also have high UV/optical luminosities comparable to J0131$-$0321 except with lower radio flux.These extremely luminous radio-loud quasars at $z>5$ will be important for exploring the powering mechanisms behind high-redshift radio-loud quasars. The origin of high UV/optical luminosities of these quasars may be due to nonthermal emission boosted by the jet, or perhaps gravitational lensing.  Since the best angular resolution for all the photometric data available is no better than $1\arcsec$, in order to investigate the origin of the extremely high luminosity, additional high-resolution images are needed.

The radio-loudness parameter $R$, defined as the ratio of the rest-frame flux densities in the radio (5 GHz) to optical (4400 \AA) bands \citep{Kellermann}, strongly correlates with the mass-accretion rate \citep{Ho2002}. Supposing that the power-law continuum of J0131-0321 between the optical and radio bands follows $f(\nu) \propto~\nu^{-0.5}$, the radio flux density at rest-frame 5 GHz is 6.9$\times$10$^{-26}$ erg s$^{-1}$cm$^{-2}$Hz$^{-1}$, the flux density at 4400 \AA~ is 6.6 $\times$ 10$^{-28}$ erg s$^{-1}$cm$^{-2}$Hz$^{-1}$. Thus, the radio-loudness of J0131$-$0321 can be derived to be $R \approx$ 100.

Recent investigations (e.g., \citealp{Momjian, Frey2010, Frey2011,Cao}) based on Very Long Baseline Interferometry (VLBI) of radio-loud quasars at high redshift  ($z>4$) have revealed a population of steep-spectrum, compact sources plausibly related to very young radio quasars associated with early galaxy formation activity. It is unclear whether J0131$-$0321 is related to this class of objects.  Additional follow-up VLBI radio observations would be helpful, as well as X-ray data that can be combined with radio and optical diagnostics (see \citealt{Wu}).

%\subsection{Formalism} \label{bozomath}

%% The equation environment wil produce a numbered display equation.

%% The \notetoeditor{TEXT} command allows the author to communicate
%% information to the copy editor.  This information will appear as a
%% footnote on the printed copy for the manuscript style file.  Nothing will
%% appear on the printed copy if the preprint or
%% preprint2 style files are used.

%% The eqnarray environment produces multi-line display math. The end of
%% each line is marked with a \\. Lines will be numbered unless the \\
%% is preceded by a \nonumber command.
%% Alignment points are marked by ampersands (&). There should be two
%% ampersands (&) per line.

%% Putting eqnarrays or equations inside the mathletters environment groups
%% the enclosed equations by letter. For instance, the eqnarray below, instead
%% of being numbered, say, (4) and (5), would be numbered (4a) and (4b).
%% LaTeX the paper and look at the output to see the results.

%% This section contains more display math examples, including unnumbered
%% equations (displaymath environment). The last paragraph includes some
%% examples of in-line math featuring a couple of the AASTeX symbol macros.

\section{ Estimate of The Black Hole Mass }

%% The displaymath environment will produce the same sort of equation as
%% the equation environment, except that the equation will not be numbered
%% by LaTeX.

Although only in very few cases has a time lag for the Mg II line been measured in reverberation mapping of AGN \citep{Reichert,Metzroth}, the line width of Mg II is known to correlate well with that of H$\beta$ in single-epoch spectra \citep{Shen2008,McGill,Wang}, suggesting that using the Mg II line can yield consistent virial mass estimates to those based on the Balmer lines. It has been further confirmed that the Mg II and Balmer lines share similar kinematics, because they deliver mutually consistent SMBH mass estimates with minimal internal scatter ($\le$0.1 dex) using the latest virial mass estimators \citep{Ho2012}. The empirical scaling relation usually adopted to estimate SMBH masses from the Mg II line and  its associated continuum at 3000~\AA\ is as follows:\\
\begin{equation}
  \log (\frac{M_{\rm BH,vir}}{M_\odot})=a+b\cdot \log (\frac{\lambda L_{\lambda}}{10^{44} {\rm erg\, s^{-1}}})+2\cdot \log (\frac{\rm FWHM}{\rm km\, s^{-1}})
\end{equation}

\begin{figure*}
\centering
 \includegraphics[width=17.0cm, height=5cm, angle=0]{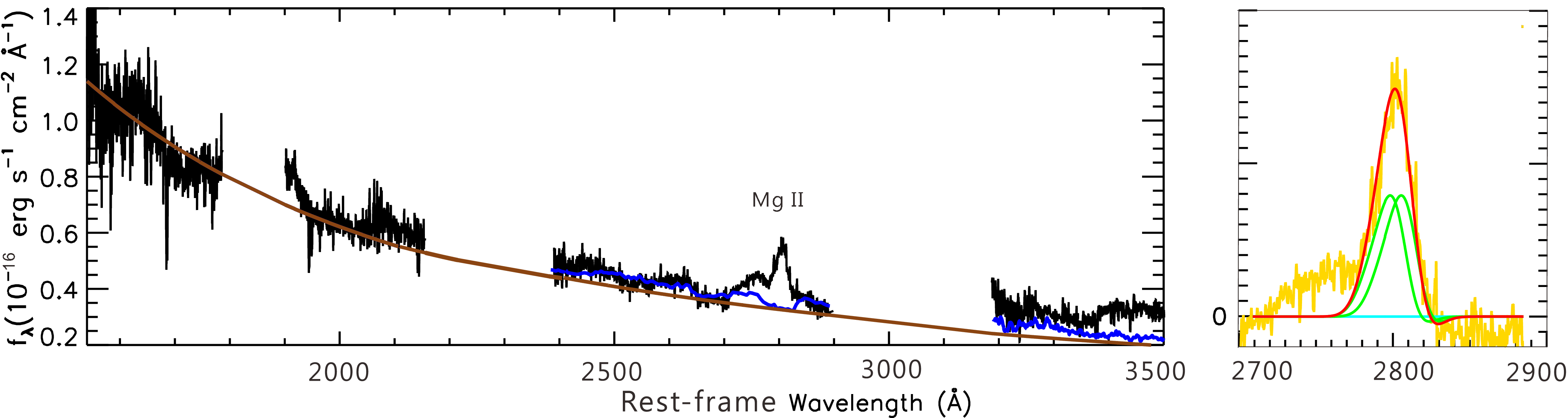}
  \caption{The rest-frame spectrum of J0131$-$0321 taken with Magellan/FIRE. The left panel shows the global fit of the pseudo-continuum, where the brown line is the power-law continuum and the blue line is the UV Fe II template. The gaps in the spectrum are due to the infrared absorption windows. The FWHM of the Mg II line  (2700$^{+100}_{-80}$ km s$^{-1}$) can be obtained through the emission-line fit with two broad Gaussian components (green lines in the right panel) after subtraction of the Fe II and continuum according to the template provided by \citet{Shen2012}. The fitted profile of the Mg II line is shown with red line in the right panel.}
  \label{fig3}
\end{figure*}
For the Mg II line, the typical values of $a$ and $b$ are 0.50 and 0.62 \citep{McLure}, respectively. This equation is calibrated from local AGNs based on broad emission-line widths and continuum luminosities. Fig. 3 presents the Magellan/FIRE near-IR spectrum of J0131$-$0321 and corresponding fitted profiles of the Mg II line, pseudo-continuum, and Fe II template. Considering the rest-frame 3000~\AA~ continuum is just red-shifted to the position of an absorption window, both global fitting and emission-line fitting are used to obtain the continuum slope and FWHM of the Mg II line. As a result, the 3000 \AA\ luminosity for J0131$-$0321 is estimated to be 2.1$^{+0.4}_{-0.3}\times$10$^{47}$ erg s$^{-1}$. Assuming an empirical conversion factor between 3000~\AA~ continuum luminosity and bolometric luminosity of $L_{\rm bol}$ = $5.18\times L_{3000}$ (e.g., \citealp{Jin,Richards}), it is easy to get the approximate bolometric luminosity of J0131$-$0321 ($L_{\rm bol} = 1.1^{+0.2}_{-0.2} \times 10^{48}$ erg s$^{-1}$), which is nearly 4.5 times greater than in the most distant quasar ULASJ1120+0641 at $z=7.085$ \citep{Mortlock}.  The BH mass is estimated to be 2.7$^{+0.5}_{-0.4} \times10^9 M_\odot$ according to the equation (1).  Assuming isotropic emission, the Eddington luminosity $L_{\rm Edd} = 1.3\times 10^{38}M_{\rm BH}/M_\odot$ erg s$^{-1}$ = $3.5 \times 10^{47}$ erg s$^{-1}$, and the Eddington ratio is $L_{\rm bol}$/$L_{\rm Edd}$ = 3.14. While in the recent research of \citet{Trakhtenbrot} , they find that the Mgii estimator may cause an underestimation of BH mass by $\sim$ 0.25 dex. Therefore, when using the latest relation, the BH mass of J0131$-$0321 is $\sim$ 4.0$\times$10$^9$ M$_\odot$, and the corresponding L$_{Edd}$ is $\sim$ 5.2$\times$10$^{47}$ erg s$^{-1}$, which means the bolometric luminosity is still larger than the Eddington luminosity. Measurement errors are small, only $\sim$ 0.1 dex.  The main uncertainties are dominated by systematic errors, which amount to at least 0.4 dex,  can be summarized as follows:

1)The scatter in the virial estimate of BH mass for the SDSS sub-sample is already $\sim 0.34$ dex \citep{Trakhtenbrot}.  We further must assume that these mass estimators can be applied to quasars at high redshifts ( $z > 5$) and high luminosities ($L_{\rm bol} > 10^{48}$ erg s$^{-1}$) cases, where they are completely untested.

2)There is evidence indicating that super-Eddington accreting BHs may exhibit a somewhat different $R_{\rm BLR} - L$ relation than sub-Eddington sources \citep{Du}.  

3)Due to limitations of the currently available data, the flux measurements are not very accurate and the assumed bolometric correction may not be suitable for this quasar.

4)We assume isotropic radiation from this quasar, while for radio-loud quasars, the measured bolometric luminosity may be biased somewhat due to nonthermal contributions and selection effects.

All the considerations mentioned above should be taken into account when calculating the BH mass and Eddington ratio. However, taken at face value, J0131$-$0321 appears to be a candidate for one of the few super-Eddington quasars known at $z>5$.


%% Included in this acknowledgments section are examples of the
%% AASTeX hypertext markup commands. Use \url without the optional [HREF]
%% argument when you want to print the url directly in the text. Otherwise,
%% use either \url or \anchor, with the HREF as the first argument and the
%% text to be printed in the second.

\section{Discussion and Conclusion}

In the subsample of radio-loud quasars at $z\gtrsim 4$, there are six objects with BH mass and Eddington ratio estimates (see Table 2) in the catalog of \citet{Shen2011}. In this table, all of the objects have relatively high Eddington ratios, and J0131$-$0321 is the only one with an Eddington ratio larger than 2 at $z>5$. 
\begin{table}[h]
\centering
 \caption{Radio-loud quasars at $z\gtrsim4$ with SMBH mass estimates from the literature}
 \begin{tabular}{lcccr}
  \hline\noalign{\smallskip}
Object Name & $z$ &  log$R$   & logM$_{BH}$   &  $L/L_{Edd}$  \\
& & & & 
\\
  \hline\noalign{\smallskip}
Archival objects & &\\
\quad SDSS J1412+0624 & 4.47 & 2.7 & 9.8   &0.20  \\ 
\quad SDSS J1420+1205 & 4.03 & 3.0 & 9.3     & 0.46                  \\
\quad SDSS J1639+4340 & 3.98 & 1.6 & 10.6     & 0.13      \\
\quad SDSS J1235$-$0003 & 4.67 & 3.0 & 9.2    & 0.25       \\
\quad CLASS J1325+1123  & 4.41 & 2.7 & 9.4    & 0.77      \\
\quad GB 1508+5714 & 4.31 & 3.8 & 8.5     & 2.93         \\
Our object & & \\
\quad SDSS J0131$-$0321  & 5.18 & 2.0 & 9.5     & 3.14         \\
  \noalign{\smallskip}\hline
\end{tabular}
\tablecomments{These error bars do not consider the inherent scatter of relation (1), which is usually given around 0.4 dex. This table is compiled from \citet{Wu}.}
\end{table}
Except for GB 1508+5714, these objects have $M_{\rm BH} > 10^9 \, M_{\odot}$. When considering spinning BHs, which have a high radiative efficiency, there is a great challenge for the formation of such large BH masses even from seeds with masses $\sim 10^3 - 10^5\, M_{\odot}$ (e.g., \citealp{Ghisellini}). At the same time, the relative dominance between the radio and optical bands seems to evolve strongly with redshift (e.g., \citealp{Singal}), which results in significant changes in the observed frequency of radio-loud quasars with redshift and UV luminosity (e.g., \citealt{Jiang2007a}). Furthermore, as the degree of radio-loudness increases with increasing radio luminosity, it would be possible to explore the connections between the efficiency of the formation of relativistic jets and accretion power, which may, in turn, depend on the combination of the evolving accretion rate and black hole spin (see \citealp{Tchekhovskoy}). It is worth noting that, apart from J0131$-$0321, the BH masses of the other six objects listed in Table 2 are all based on the C IV emission line, which could be highly uncertain due to nonvirial kinematics in the high-ionization C IV line region \citep{Shen2012}. The BH mass of J0131$-$0321 is based on Mg II, and thus should be more reliable.  For radio-loud sources in general, if we cannot eliminate jet contamination, use of the size - continuum luminosity relation might overestimate the size of the broad-line region (BLR) size and hence the BH mass \citep{wxb2004}. The broad emission lines, however, can be seen clearly in all spectra, which suggests that the beamed relativistic jets may have a small contribution to the UV/optical continua in these sources.

The observational data for J0131$-$0321 are currently very limited.  It would be particularly valuable to do multi-wavelength follow-up observations with more dedicated and advanced facilities, such as VLBI for the radio regime and {\it Chandra}\ or {\it XMM-Newton}\ for X-ray energies. These future observations will enable us to investigate further the evolution of radio-loud quasars and to explore their powering mechanisms, which are still unclear.

\acknowledgments

We acknowledge the support of the staff of the LJT. Funding for the LJT has been provided by CAS and the People's Government of Yunnan Province. XBW thanks the supports by the NSFC grants 11033001 and 11373008. JMB is supported by the NSFC grants 11133006 and 11361140347. XBW,JMB and LCH are all supported by the Strategic Priority Research Program \lq\lq The Emergence of Cosmological Structures \rq\rq of the Chinese Academy of Sciences (grant No. XDB09000000). WNB thanks support from NASA ADP grant NAG5-13035. This paper includes data gathered with the 6.5 meter Magellan Telescopes located at Las Campanas Observatory, Chile. The MagE spectral reduction was kindly provided by George Becker. We also acknowledge the use of the quasar spectral fitting code kindly provided by Yue Shen.

\end{document}